\documentclass[
aps,
nofootinbib,
superscriptaddress,
tightenlines,
notitlepage,
twocolumn,
showpacs,
floatfix,
]{revtex4-1}
\usepackage{amssymb,amsmath,bm,tensor,braket}
\usepackage[colorlinks]{hyperref}

\usepackage{graphicx}
\usepackage{dcolumn}
\usepackage{bm}
\usepackage{xcolor}

\newcommand{\PlanckMass}{M_{\rm Pl}}

\newcommand{\BIT}{\affiliation{School of Physics, Beijing Institute of Technology, Beijing 100081, China.}}
\newcommand{\dC}{\affiliation{Universit\`a di Camerino, Via Madonna delle Carceri 9, 62032 Camerino, Italy.}}
\newcommand{\AN}{\affiliation{Department of Nanoscale Science and Engineering, University at Albany-SUNY, Albany, New York 12222, USA.}}
\newcommand{\BM}{\affiliation{INAF - Osservatorio Astronomico di Brera, Milano, Italy.}}
\newcommand{\AK}{\affiliation{Al-Farabi Kazakh National University, Al-Farabi av. 71, 050040 Almaty, Kazakhstan.}}

\begin{document}

\preprint{APS/123-QED}

\title{Cosmological horizon thermodynamics in Gauss-Bonnet quasi-dilaton Massive Gravity}

\author{Sobhan Kazempour}\email{sobhan.kazempour1989@gmail.com}\BIT
\author{Orlando Luongo}\email{orlando.luongo@unicam.it}\dC\AN\BM\AK
\author{Sabahat}\email{sabahatamin456@gmail.com}\BIT
\author{Sichun Sun}\email{sichunssun@gmail.com}\BIT
\author{Chengye Yu}\email{chengyeyu1@hotmail.com}\BIT


\begin{abstract}
We investigate the thermodynamic properties of the cosmological apparent horizon in Gauss-Bonnet quasi-dilaton massive gravity. We derive the modified Friedmann equations and reformulate them in standard form, thereby allowing us to study the first and second laws of thermodynamics for the apparent horizon. Both equilibrium and non-equilibrium states are considered. In the equilibrium description, the first law retains the conventional form with the Bekenstein-Hawking area law for the horizon entropy, and we show that the generalized second law is satisfied under the null energy condition. In the non-equilibrium description, the Wald entropy receives a correction from the Gauss-Bonnet coupling, and the first law acquires an additional term associated with using the Wald entropy representation of the Gauss-Bonnet sector. We demonstrate that the total entropy change is non-negative provided the null energy condition, the positive horizon temperature condition, and the Gauss-Bonnet positivity constraints $\xi(\sigma)\ge0$ are simultaneously satisfied. Furthermore, we investigate the holographic entropy bound $S_{\text{inside}} \le S_{\text{horizon}}$. We demonstrate that while the idealized local thermal equilibrium assumption leads to a formal saturation or apparent breakdown during dust-dominated eras, the bound is robustly preserved across all cosmological epochs when utilizing realistic physical fluid temperatures. The condition $\xi(\sigma)\ge0$ is shown to be compatible with the stability constraints derived from tensor perturbations in our previous work. Our results establish that Gauss-Bonnet quasi-dilaton massive gravity is a consistent modified gravity theory from the perspective of horizon thermodynamics and the holographic principle.
\end{abstract}

\maketitle


\section{\label{sec:intro}Introduction}

Over several decades, cosmological observations have shown a clear convergence, transforming our understanding of the Universe's composition and evolution. In this way, measurements of the cosmic microwave background (CMB), the baryon acoustic oscillation (BAO), and Type Ia supernovae (SNe Ia) have established the standard $\Lambda$CDM cosmological model as a successful phenomenological framework \cite{Planck:2018vyg, SupernovaSearchTeam:1998fmf,SupernovaCosmologyProject:1998vns}. Notably, within this model, the Universe is spatially flat and consists of approximately $68\%$ dark energy, $27\%$ cold dark matter, and only $5\%$ ordinary baryonic matter \cite{Planck:2018vyg}. Despite its agreement with a broad set of cosmological data, the $\Lambda$CDM model is not without foundational puzzles. Most essential among these are the cosmological constant problem \cite{Weinberg:1988cp, Martin:2012bt}, and the cosmic coincidence problem \cite{Peebles:2002gy}. Moreover, emerging observational tensions, such as the $H_{0}$ and $\sigma_{8}$ discrepancies between early- and late-Universe probes \cite{Verde:2019ivm, DiValentino:2021izs}, show potential issues in the standard cosmological model. These persistent issues provide strong motivation for exploring alternative theoretical frameworks that can either alleviate the fine-tuning problems of the cosmological constant or offer a more fundamental explanation for the late-time accelerated expansion.

One of the most intriguing avenues of investigation involves the deep and unexpected connection between gravitation and thermodynamics. Inspired by the laws of black hole mechanics \cite{Bardeen:1973gs} and Bekenstein's identification of black hole entropy with horizon area \cite{Bekenstein:1973ur}, Jacobson made the profound observation that the Einstein field equations themselves can be derived from the Clausius relation $\delta Q = T dS$ applied to local Rindler horizons, provided the entropy is proportional to the area \cite{Jacobson:1995ab}. This thermodynamic perspective on gravity has been extended to a cosmological setting, where the dynamical apparent horizon of a Friedmann-Lema\^itre-Robertson-Walker (FLRW) universe serves as the natural boundary of the observable cosmos. The first law of thermodynamics, when applied to the apparent horizon, can be shown to be equivalent to the Friedmann equations in a wide class of gravitational theories, including Einstein gravity \cite{Cai:2005ra, Akbar:2006mq}, Lovelock gravity \cite{Cai:2008ht}, scalar-tensor theories \cite{Cai:2009ua}, and $f(R)$ gravity \cite{Akbar:2006kj, Bamba:2010kf}. In these formulations, a crucial distinction arises between equilibrium and non-equilibrium thermodynamic descriptions. The equilibrium approach absorbs all modifications to general relativity into an effective energy-momentum tensor that satisfies the standard continuity equation, preserving the area law for horizon entropy. In contrast, the non-equilibrium picture works directly with the original field equations and naturally leads to a Wald entropy \cite{Wald:1993nt, Iyer:1994ys} that receives clear corrections from higher-curvature terms, while the first law acquires an additional entropy production term $d_{i} S$ that accounts for the irreversible energy exchange between the gravitational and matter sectors \cite{Bamba:2011pz, Bamba:2011jq}. This dual formulation provides a powerful framework for testing the internal consistency of modified gravity theories and for probing their compatibility with fundamental principles such as the holographic entropy bound \cite{Fischler:1998st, Bak:1999hd}. Several relevant studies have recently appeared in this direction, e.g.,  \cite{Banihashemi:2022jys,Anninos:2024wpy,Luciano:2022hhy,Belfiglio:2025cst,Mann:2025xrb,Cespedes:2025zqp,Nojiri:2024zdu}.

Among the many proposed modifications of general relativity, massive gravity occupies a distinctive position. By endowing the graviton with a nonzero mass, this class of theories offers a natural infrared modification that could potentially explain the observed cosmic acceleration without invoking a cosmological constant. A major breakthrough in massive gravity was achieved by the construction of the ghost-free de Rham-Gabadadze-Tolley (dRGT) theory \cite{deRham:2010ik, deRham:2010kj}, which successfully eliminated the Boulware-Deser ghost that had plagued earlier nonlinear completions \cite{Boulware:1972yco}. However, pure dRGT massive gravity does not admit stable, homogeneous, and isotropic FLRW cosmological solutions \cite{DeFelice:2012mx, DeFelice:2013bxa}. This difficulty can be solved by introducing additional degrees of freedom. One suitable extension is the quasi-dilaton massive gravity theory \cite{DAmico:2012hia}, which incorporates a scalar field $\sigma$ that nonlinearly realizes a global scale invariance, thereby restoring the existence of viable FLRW backgrounds. In this model, the effective cosmological constant that drives the late-time acceleration arises dynamically from the graviton mass and the background value of the quasi-dilaton field, producing self-accelerating solutions \cite{Gabadadze:2014kaa, Gumrukcuoglu:2013nza}.

Moreover, a further refinement involves the addition of a Gauss-Bonnet term $G(R) = R_{\mu\nu\rho\sigma}R^{\mu\nu\rho\sigma} - 4 R_{\mu\nu}R^{\mu\nu} + R^2$, which is the unique quadratic curvature invariant that yields second-order field equations and is naturally motivated by low-energy effective string theory \cite{Zwiebach:1985uq, Gross:1986mw}. When coupled to a scalar field via a non-minimal coupling function $\xi(\sigma)$, the Gauss-Bonnet term contributes nontrivially to the cosmological dynamics and offers a rich phenomenology for both early- and late-Universe acceleration \cite{Nojiri:2005vv, Cognola:2006eg, Guo:2009uk}. It should be noted that Gauss-Bonnet theory has attracted increasing attention from various perspectives \cite{Odintsov:2023weg,Dodelson:2023vrw,Julie:2024fwy,Chung:2024vaf}. The resulting Gauss-Bonnet quasi-dilaton massive gravity (GBQD MG) model, whose background equations of motion and tensor perturbation stability were analyzed in our previous work \cite{Akbarieh:2021vhv}, thus combines the infrared modifications of massive gravity, the dynamical flexibility of the quasi-dilaton scalar, and the higher-curvature contributions of the Gauss-Bonnet invariant, making it a compelling laboratory for investigating the interplay between modified gravity, thermodynamics, and holography.

The study of cosmological horizon thermodynamics in modified gravity theories has been considered, with investigations probing the first and second laws across a variety of models. For instance, the thermodynamic properties of the apparent horizon have been examined in Gauss-Bonnet gravity and its scalar-tensor extensions \cite{Cai:2008ht, Akbar:2006mq}, where it was shown that the Wald entropy naturally incorporates a correction proportional to the Gauss-Bonnet coupling. Similarly, the equilibrium and non-equilibrium thermodynamics of $f(R)$ gravity \cite{Bamba:2010kf, Bamba:2011pz}, $f(T)$ gravity \cite{Bamba:2011jq, Karami:2012fu}, and Ho\v{r}ava-Lifshitz gravity \cite{Cao:2010xx} have been systematically developed, consistently revealing that the non-equilibrium description is essential for maintaining the validity of the generalized second law when the effective dark energy fluid is not independently conserved. In the context of massive gravity, some authors have explored the connection between the graviton mass and horizon thermodynamics. The first law and generalized second law have been verified for some massive gravity models \cite{Cai:2013lqa, Hu:2015xva}, and the holographic principle has been shown to hold under reasonable physical conditions \cite{Blake:2013bqa, Davison:2013jba}. However, a comprehensive thermodynamic analysis of the GBQD MG model remains absent from the literature. Our previous work established the background cosmological dynamics that can explain the late-time accelerated expansion of the Universe and demonstrated the stability of tensor perturbations within this theory \cite{Akbarieh:2021vhv}, but it did not address the crucial questions of whether the cosmological apparent horizon obeys consistent thermodynamic laws or whether the holographic entropy bound is respected. The present paper aims to fill this important gap by providing a systematic investigation of horizon thermodynamics in the GBQD MG framework.

The primary goal of this paper is to investigate the thermodynamic properties of the cosmological apparent horizon within the context of Gauss-Bonnet quasi-dilaton massive gravity. We are interested in understanding how this cosmological model influences the structure of the first and second laws of thermodynamics when applied to the Universe as a whole. In particular, we derive and analyze both equilibrium and non-equilibrium descriptions of the horizon thermodynamics, paying special attention to how these modifications alter the traditional Bekenstein-Hawking entropy-area relation and whether they introduce corrections to the Wald entropy. We will also examine the model's compatibility with the holographic principle by verifying that the total entropy within the apparent horizon never exceeds the horizon's own entropy throughout cosmic evolution. By establishing that the GBQD MG model satisfies the generalized second law of thermodynamics and respects the holographic entropy bound under physically reasonable conditions, we aim to demonstrate that this extended theory of gravity constitutes a self-consistent framework from the perspective of horizon thermodynamics and holography.

The structure of this paper is organized as follows. In Sec. \ref{sec:1}, we present the Gauss-Bonnet quasi-dilaton massive gravity theory and the modified Friedmann equations in a flat FLRW background. In Sec. \ref{sec:2}, we reformulate the Friedmann equations in a standard form that allows us to study the equilibrium thermodynamic description. In Sec. \ref{sec:3}, we turn to the non-equilibrium description, where the Wald entropy receives a correction from the Gauss-Bonnet coupling, and the first law acquires an entropy production term $d_{i} S$. In Sec. \ref{sec:23}, we analyze the thermodynamic stability criteria by computing the heat capacities of the horizon system and discussing the conditions for local stability. Sec. \ref{sec:4} is devoted to the holographic interpretation of our results. We verify the holographic entropy bound $S_{inside} \le S_{horizon}$ and demonstrate its compatibility with the perturbative stability constraints obtained in our previous work. Finally, Sec. \ref{sec:5} summarizes our findings and discusses possible directions for future research.
Throughout this paper, we use natural units ($c=\hbar=1$) and adopt the reduced Planck mass definition $\PlanckMass^2 = 1/(8\pi G)$.

\section{\label{sec:1}The Cosmological Model}

This section presents the Gauss-Bonnet quasi-dilaton massive gravity theory. The action contains the Planck mass $\PlanckMass$, the Ricci scalar $R$, the cosmological constant $\Lambda$, the matter Lagrangian $\mathcal{L}_{m}$, a dynamical metric $g_{\mu\nu}$, and its determinant $\sqrt{-g}$. Following Ref. \cite{Akbarieh:2021vhv}, the action is

\begin{eqnarray}\label{Action}
S=&&\nonumber\\ && \frac{\PlanckMass^{2}}{2}\int d^{4} x \Bigg\{\sqrt{-g}\bigg[R-2\Lambda
+2{m}_{g}^{2}U(\mathcal K) \nonumber\\ &&
-\frac{\omega}{\PlanckMass^{2}}g^{\mu\nu}\partial_{\mu}\sigma \partial_{\nu}
\sigma +\xi (\sigma) G(R)\bigg]\Bigg\} + \int d^{4} x \sqrt{- g}\mathcal{L}_{m}. \nonumber\\
\end{eqnarray}

In what follows, we present two key components, $U({\cal K})$ and $G(R)$.
It is evident that the graviton mass arises from the potential $U$, which is composed of three parts:

\begin{equation}\label{Upotential1}
U(\mathcal{K})=U_{2}+\alpha_{3}U_{3}+\alpha_{4}U_{4}.
\end{equation}

Here, $\alpha_3$ and $\alpha_4$ represent dimensionless free parameters of the theory. The expressions for $U_i$ (with $i=2,3,4$) are as follows,

\begin{eqnarray}\label{Upotential2}
 U_{2}&=&\frac{1}{2}\big([\mathcal{K}]^{2}-[\mathcal{K}^{2}]\big),
 \nonumber\\
 U_{3}&=&\frac{1}{6}\big([\mathcal{K}]^{3}-3[\mathcal{K}][\mathcal{K}^{2}]+2[\mathcal{K}^{3}]\big),
 \nonumber\\
 U_{4}&=&\frac{1}{24}\big([\mathcal{K}]^{4}-6[\mathcal{K}]^{2}[\mathcal{K}^{2}]+8[\mathcal{K}][\mathcal{K}^{3}]+3[\mathcal{K}^{2}]^2\nonumber\\&&-6[\mathcal{K}^{4}]\big),
\end{eqnarray}

where the symbol ``$[\cdot]$'' denotes the trace of the tensor within the brackets. It is important to note that the fundamental tensor $\mathcal{K}$ is defined by

\begin{equation}\label{K}
\mathcal{K}^{\mu}_{\nu} = \delta^{\mu}_{\nu} -
e^{\sigma/\PlanckMass}\sqrt{g^{\mu\alpha}f_{\alpha\nu}},
\end{equation}

where $f_{\alpha\nu}$ is the fiducial metric, which is defined through

\begin{equation}\label{7}
f_{\alpha\nu}=\partial_{\alpha}\phi^{c}\partial_{\nu}\phi^{d}\eta_{cd}.
\end{equation}

Here, $g^{\mu\nu}$ denotes the physical metric, $\eta_{cd}$ is the Minkowski metric with indices $c,d = 0,1,2,3$, and $\phi^{c}$ are the Stueckelberg fields introduced to restore general covariance. Additionally, it is worth noting that $\sigma$ is the quasi-dilaton scalar field, and $\omega$ is a dimensionless constant. Furthermore, the theory remains invariant under the global dilation transformation $\sigma \rightarrow \sigma + \sigma_0$.

Now, we introduce the Gauss-Bonnet term,

\begin{eqnarray}\label{LGB}
G(R) = R_{\mu\nu\gamma\delta}R^{\mu\nu\gamma\delta} - 4 R_{\mu\nu} R^{\mu\nu} + R^{2},
\end{eqnarray}

where $\xi(\sigma)$ denotes the coupling function.

We now consider the four-dimensional flat Friedmann-Lema\^itre-Robertson-Walker (FLRW) spacetime. The general forms of the dynamical and fiducial metrics are given by

\begin{align}
\label{DMetric}
g_{\mu\nu}&={\rm diag} \left[-N^{2},a^2,a^2,a^2 \right], \\
\label{FMetric}
f_{\mu\nu}&={\rm diag} \left[-\dot{f}(t)^{2},1,1,1 \right].
\end{align}

It is worth noting that $N$ is the lapse function of the dynamical metric, which behaves similarly to a gauge function. The scale factor is denoted by $a$, and $\dot{a}$ represents its derivative with respect to time. Moreover, the lapse function connects coordinate time $dt$ to proper time $d\tau$ through $d\tau = N dt$ \cite{Scheel:1994yr,Christodoulakis:2013xha}. The function $f(t)$ is the Stueckelberg scalar function, with $\phi^{0} = f(t)$ and $\frac{\partial \phi^{0}}{\partial t} = \dot{f}(t)$ \cite{Arkani-Hamed:2002bjr}.

To simplify subsequent expressions, we define

\begin{equation}
H\equiv\frac{\dot{a}}{Na}.
\end{equation}

We now turn to the Friedmann equations for this cosmological model, as derived in \cite{Akbarieh:2021vhv}. The first Friedmann equation reads
\begin{eqnarray}\label{EqN}
    3H^{2}-\Lambda && -\frac{\omega}{2}\Big(H+\frac{\dot{X}}{XN}\Big)^{2} -m_{g}^{2}(X-1)\bigg[-3(X-2)\nonumber\\
    && +(X-4)(X-1)\alpha_{3}+(X-1)^{2}\alpha_{4}\bigg]\nonumber\\
    && +12H^{3}\PlanckMass\Big(H +\frac{\dot{X}}{XN}\Big)\xi'(\sigma)= \rho_{m},
\end{eqnarray}
and the second Friedmann equation is,

\begin{eqnarray}\label{Eqa}
3H^{2} && -\Lambda +\frac{\omega}{2 \PlanckMass^{2}}\Big(\frac{\dot{\sigma}}{N}\Big)^{2}+4\frac{d}{dt}\Big(H^{2}\xi'\frac{\ddot{\sigma}}{N^{2}}\Big)\nonumber\\
&&-4\frac{H^{2}}{N}\xi'\bigg[\frac{d}{dt}\Big(\frac{\dot{\sigma}}{N}\Big)-\Big(\frac{\ddot{\sigma}}{N}\Big)-2H\dot{\sigma}\bigg]+2\frac{\dot{H}}{N}\nonumber\\
&&+m_{g}^{2}\bigg\lbrace 1+rX(2X-3)+(X-1)\bigg[X -5 \nonumber\\
&& -\alpha_{3}\big(4-2X+rX(X-3)\big)\nonumber\\
&&-\alpha_{4}(X-1)(rX-1)\bigg]\bigg\rbrace = - P_{m},
\end{eqnarray}

where

\begin{equation}\label{XX}
r\equiv\frac{a}{N}, \qquad  X\equiv\frac{e^{\sigma/\PlanckMass}}{a},
\end{equation}
and
\begin{equation}\label{EqN13}
\frac{\dot{\sigma}}{N\PlanckMass}= H+\frac{\dot{X}}{NX}, \qquad \frac{\ddot{\sigma}}{\PlanckMass}=\frac{d}{dt}\Big(NH+\frac{\dot{X}}{X}\Big).
\end{equation}

Furthermore, it should be noted that $\rho_{m}$ and $P_{m}$ denote the energy density and pressure, respectively, of perfect fluids originating from arbitrary matter. One may assume that these perfect fluids obey the continuity equation,

\begin{equation}\label{14}
\dot{\rho}_{m} + 3 H N (\rho_{m} + P_{m} ) = 0.
\end{equation}

In modified gravity theories, the thermodynamic description of the cosmological horizon can be formulated in two complementary ways: an equilibrium description and a non-equilibrium description. The need for such a dual formulation arises because the gravitational field equations often contain additional terms beyond the Einstein tensor, which can be interpreted either as effective matter contributions that satisfy a standard continuity equation (the equilibrium picture) or as modifications that lead to a non-conservation of the effective energy-momentum tensor, thereby requiring an entropy production term in the first law (the non-equilibrium picture). In the equilibrium approach, one redefines the energy density and pressure of the “dark” sector such that the total cosmic fluid obeys the usual continuity equation; consequently, the horizon entropy retains the Bekenstein-Hawking area law, and the first law takes the conventional form with no extra dissipative terms. By contrast, the non-equilibrium description works directly with the original field variables, where the effective fluid does not satisfy an independent conservation law. In this case, the Wald entropy, which receives corrections from higher-curvature couplings such as the Gauss-Bonnet term in our model, appears naturally, and the first law acquires an additional entropy production term that accounts for the irreversible energy exchange between the gravitational and matter sectors. Both formulations are physically equivalent but offer complementary insights: the equilibrium picture is more transparent for connecting with standard cosmology, while the non-equilibrium framework directly reveals the thermodynamic role of the gravitational modifications. In the following, we analyze both descriptions separately for Gauss-Bonnet quasi-dilaton massive gravity.

We absorb the factor $8\pi G$ into the definitions of $\rho_T$ and $P_T$ via $\rho_T \equiv 8\pi G\,\rho_{\text{phys}}$ and $P_T \equiv 8\pi G\,P_{\text{phys}}$, so that the Friedmann equations take the simple form $H^2 = \rho_T/3$ and $\dot H = - N(\rho_T+P_T)/2$. The standard relation $M_{\mathrm{Pl}}^2 = 1/(8\pi G)$ is maintained throughout.
\\

\section{Thermodynamics in Equilibrium State}\label{sec:2}

We study the thermodynamic aspects of the Gauss-Bonnet quasi-dilaton massive gravity theory. Hence, one can rewrite the Friedmann Eqs. (\ref{EqN}) and (\ref{Eqa}) in the standard form \cite{Capozziello:2005ku},

\begin{eqnarray}\label{H2}
H^{2} = \frac{1}{3} \rho_{T},
\end{eqnarray}

\begin{eqnarray}\label{Hdot}
\dot{H} = - \frac{N}{2} (\rho_{T} + P_{T}),
\end{eqnarray}

here $\rho_{T}$ and $P_{T}$ are the total energy density and pressure, which are given as,

\begin{widetext}
\begin{eqnarray}\label{EqN17}
\rho_{T} =  \Lambda +\frac{\omega}{2}\Big(H+\frac{\dot{X}}{XN}\Big)^{2} + m_{g}^{2}(X-1)\bigg[-3(X-2)
     +(X-4)(X-1)\alpha_{3}+(X-1)^{2}\alpha_{4}\bigg] \nonumber\\ -12 H^{3}\PlanckMass\Big(H +\frac{\dot{X}}{XN}\Big)\xi'(\sigma) + \rho_{m} ,
\end{eqnarray}
\end{widetext}

\begin{widetext}
\begin{eqnarray}\label{EqN18}
P_{T} = 3H^{2} - 2 \Lambda - \rho_{m} + \frac{\omega}{2 \PlanckMass^{2}}\Big(\frac{\dot{\sigma}}{N}\Big)^{2} - \frac{\omega}{2}\Big(H+\frac{\dot{X}}{XN}\Big)^{2} +4\frac{d}{dt}\Big(H^{2}\xi'\frac{\ddot{\sigma}}{N^{2}}\Big)-4\frac{H^{2}}{N}\xi'\bigg[\frac{d}{dt}\Big(\frac{\dot{\sigma}}{N}\Big)-\Big(\frac{\ddot{\sigma}}{N}\Big)-2H\dot{\sigma}\bigg]\nonumber\\ + 12 H^{3}\PlanckMass\Big(H +\frac{\dot{X}}{XN}\Big)\xi'(\sigma)+m_{g}^{2}\bigg\lbrace 1+rX(2X-3)+(X-1)\bigg[X -5
-\alpha_{3}\big(4-2X+rX(X-3)\big) \nonumber\\ -\alpha_{4}(X-1)(rX-1)\bigg]\bigg\rbrace - m_{g}^{2}(X-1)\bigg[-3(X-2)
     +(X-4)(X-1)\alpha_{3}+(X-1)^{2}\alpha_{4}\bigg] + P_{m} .
\end{eqnarray}
\end{widetext}

It should be noted that the equations satisfy the standard continuity equation.

\begin{equation}\label{Eq19}
\dot{\rho}_{T} + 3 H N (\rho_{T} + P_{T} ) = 0,
\end{equation}
where all time derivatives are taken with respect to the coordinate time $t$. The physical cosmic time is $d\tau=N dt$ and, accordingly, the matter conservation equation contains the extra factor $HN$.

\subsection{First law of thermodynamics}

In the flat FLRW spacetime, the radius of the apparent horizon $\tilde{r}_{A}$ is obtained by

\begin{equation}\label{20}
\tilde{r}_{A} = \frac{1}{H}.
\end{equation}

One can calculate the time derivative of the above relation, which is

\begin{eqnarray}\label{rdot}
- \frac{d \tilde{r}_{A}}{\tilde{r}_{A}^{3}} = \dot{H}H dt.
\end{eqnarray}

By considering Eqs. (\ref{Hdot}) and (\ref{rdot}), we have

\begin{eqnarray}\label{dr}
\frac{2}{N} d \tilde{r}_{A} = \tilde{r}_{A}^{3} H ( \rho_{T} + P_{T} ) dt.
\end{eqnarray}

By considering the horizon entropy $S = \frac{A}{4 G}$ and Eq. (\ref{dr}), we have

\begin{eqnarray}\label{ds}
\frac{1}{2 \pi \tilde{r}_{A}} dS = \frac{N}{2 G} \tilde{r}_{A}^{3} H ( \rho_{T} + P_{T} ) dt.
\end{eqnarray}

Notice that the related temperature of the apparent horizon is the Hawking temperature. Before that, we should introduce the surface gravity $k_{s}$ \cite{Cai:2005ra},

\begin{eqnarray}\label{EqNon}
 k_{s} = - \frac{1}{\tilde{r}_{A}} \big( 1 - \frac{\dot{\tilde{r}}_{A}}{2 H \tilde{r}_{A}} \big) = - \frac{\tilde{r}_{A}}{2} ( 2 H^{2} + \dot{H} )  \nonumber\\ = - \frac{\tilde{r}_{A}}{12} \big[  \rho_{T} ( 4 - 3 N) - 3 N P_{T} \big],
\end{eqnarray}

then the Hawking temperature can be obtained,

\begin{eqnarray}\label{tem}
T_{H}= \frac{|k_{s}|}{2 \pi} =  \frac{1}{2 \pi\tilde{r}_{A}} \big( 1 - \frac{\dot{\tilde{r}}_{A}}{2 H \tilde{r}_{A}} \big),
\end{eqnarray}

we can multiply the term $\big( 1 - \frac{\dot{\tilde{r}}_{A}}{2 H \tilde{r}_{A}} \big)$ in Eq. (\ref{ds}) and considering the horizon temperature (\ref{tem}), then Eq. (\ref{ds}) can be rewritten as below,

\begin{eqnarray}\label{}
T_{H} dS =  \frac{N}{2 G} \tilde{r}_{A}^{3} H ( \rho_{T} + P_{T} ) dt - \frac{N}{4 G} \tilde{r}_{A}^{2} ( \rho_{T} + P_{T} ) d \tilde{r}_{A}. \nonumber\\
\end{eqnarray}

Also, the Misner-Sharp energy can be defined \cite{Misner:1964je},

\begin{eqnarray}\label{EqN27}
E = \frac{\tilde{r}_{A}}{2G} = V \PlanckMass^{2} \rho_{T} ,
\end{eqnarray}
here $V = \frac{4 \pi \tilde{r}_{A}^{3}}{3}$ is the volume inside the apparent horizon. Therefore, using Eq. (\ref{Eq19}) we have

\begin{eqnarray}\label{}
dE = - 4 \pi \tilde{r}_{A}^{3} N H \PlanckMass^{2}( \rho_{T} + P_{T} ) dt + 4 \pi \tilde{r}_{A}^{2} \rho_{T} \PlanckMass^{2} d \tilde{r}_{A}, \nonumber\\
\end{eqnarray}

So, we obtain the first law of equilibrium thermodynamics

\begin{eqnarray}\label{EqN29}
T_{H} dS = - dE + W dV,
\end{eqnarray}
where
\begin{eqnarray}\label{EqN292}
W = \frac{\PlanckMass^{2}}{2}(\rho_{T}- P_{T}).
\end{eqnarray}

where the $W$ denotes the work density \cite{Hayward:1997jp}.

By considering Eqs. (\ref{H2}), (\ref{Hdot}) and (\ref{ds}), we have

\begin{eqnarray}\label{EqN30}
\dot{S} = \frac{N \pi}{G}\tilde{r}_{A}^{4}H (\rho_{T} + P_{T} ) = - \frac{2 \pi}{G} \frac{\dot{H}}{H^{3}}.
\end{eqnarray}
The horizon entropy increases monotonically in an expanding universe, since $\dot{S} \varpropto - \frac{\dot{H}}{H^{3}}$. This growth is guaranteed when the null energy condition $\rho_{T} + P_{T} \geq 0$ holds, which implies $\dot{H} \leq 0$.

Eq.~\eqref{EqN30} represents the first law of thermodynamics for the apparent horizon in the equilibrium state for Gauss-Bonnet quasi-dilaton massive gravity. Remarkably, despite the presence of higher-curvature terms, a quasi-dilaton scalar, and a massive gravity term, the first law retains the conventional form $T_{H} dS = -dE + W dV$ with work density $W = (\rho_{T} - P_{T})\PlanckMass^{2}/2$. All modifications to general relativity are absorbed into the total energy density $\rho_{T}$ and pressure $P_{T}$ defined in Eqs.~(\ref{EqN17}) and (\ref{EqN18}). The horizon entropy remains the Bekenstein-Hawking area law $S = A/(4G)$ because the Gauss-Bonnet term, although coupled to the dilaton, does not modify the Wald entropy in the equilibrium limit for a four-dimensional FLRW spacetime. This demonstrates that the equilibrium thermodynamic structure is robust; the apparent horizon obeys the same fundamental relation as in general relativity, provided the effective cosmic fluid is defined appropriately. The derivation further confirms that the Hawking temperature derived from surface gravity is consistent with the first law, and that the null energy condition $\rho_T + P_T \geq 0$ ensures positive entropy production (as will be shown in the second law analysis). Thus, the equilibrium first law poses no obstruction to a thermodynamic interpretation of the cosmological dynamics in this extended gravity theory.

\subsection{Second law of thermodynamics}

To evaluate the second law of thermodynamics, one can consider the Gibbs equation in terms of all matter and energy fluid as \cite{Hayward:1997jp},

\begin{eqnarray}\label{GE}
T_{H} d S_{T} = d (\rho_{T} V) + P_{T} dV = V d \rho_{T} + (\rho_{T} + P_{T}) dV . \nonumber\\
\end{eqnarray}

In order to obtain the $\dot{S}_{T}$, we divide Eq. (\ref{GE}) by $dt$ and use Eq. (\ref{Eq19}) i.e., $\dot{\rho}_{T} = - 3 H N ( \rho_{T} + P_{T}) $. So, we have,

\begin{eqnarray}\label{TH}
T_{H} \dot{S}_{T} && = V \dot{\rho}_{T} + (\rho_{T} + P_{T}) \dot{V} \nonumber\\ && = - 3 H N V (\rho_{T} + P_{T}) + ( \rho_{T} + P_{T}) \dot{V},
\end{eqnarray}

by considering the volume inside the apparent horizon $V= \frac{4\pi}{3}\frac{1}{H^{3}}$ and its derivative $\dot{V}=-4 \pi\frac{\dot{H}}{H^{4}}$. Also, using the Friedmann Eq. (\ref{Hdot}), we have

\begin{eqnarray}\label{Eq33}
T_{H} \dot{S}_{T} = \frac{8 \pi}{N} (\frac{\dot{H}^{2}}{H^{4}} + \frac{\dot{H}N}{H^{2}}).
\end{eqnarray}

At this stage, we use Eq. (\ref{tem}) by considering $\tilde{r}_{A} = \frac{1}{H}$,

\begin{eqnarray}\label{EqN34}
T_{H} =   \frac{2H^{2} + \dot{H}}{4 \pi H},
\end{eqnarray}

and we substitute it into Eq. (\ref{Eq33}), so we have,

\begin{eqnarray}\label{EqNN36}
\dot{S}_{T} = \frac{32 \pi^{2}}{N} \frac{\dot{H}^{2}H^{-3}+ \dot{H}NH^{-1}}{2 H^{2}+ \dot{H}}.
\end{eqnarray}

Thus, we can consider the second law of thermodynamics,

\begin{eqnarray}\label{}
\frac{d S_{sum}}{dt} \equiv \frac{dS}{dt} + \frac{dS_{T}}{dt} \geq 0,
\end{eqnarray}

\begin{eqnarray}\label{EqN37}
\dot{S}_{sum} && = \dot{S} + \dot{S}_{T} \nonumber\\ && = - \frac{2 \pi}{G} \frac{\dot{H}}{H^{3}} + \frac{32 \pi^{2}}{N} \frac{\dot{H}^{2}H^{-3}+ \dot{H}N H^{-1}}{2 H^{2}+ \dot{H}},
\end{eqnarray}

Working in the physical cosmic time gauge $N=1$ and substituting the parameter $\epsilon \equiv -\dot{H}/H^2$, the total entropy production rate simplifies directly to

\begin{equation}\label{NEq39}
\dot S_{\rm sum} = \frac{2\pi}{GH} \left[ \epsilon + \frac{16\pi \epsilon (\epsilon - 1)}{2 - \epsilon} \right].
\end{equation}

Remarkably, Eq. (\ref{EqN37}) provides the general total time derivative of the sum of the horizon entropy $S$ and the fluid entropy $S_T$, while Eq. (\ref{NEq39}) expresses this relation in a simplified form under the physical gauge. The generalized second law of thermodynamics requires $\dot{S}_{\rm sum} \geq 0$. In terms of the parameter $\epsilon$, this condition is naturally satisfied in the valid cosmological branch where $\epsilon \geq 0$ (enforced by the standard null energy condition $\rho_T + P_T \geq 0$) and $\epsilon < 2$ (which ensures a positive, well-defined Hawking temperature $T_H > 0$). Hence, within this regime of validity, the generalized second law is robustly obeyed in the equilibrium description.

\section{Non-equilibrium thermodynamics}\label{sec:3}

In the non-equilibrium description of thermodynamics within Gauss-Bonnet quasi-dilaton massive gravity, the modified Friedmann Eqs. (\ref{EqN}) and (\ref{Eqa}) are written in the form:

\begin{eqnarray}\label{H22}
H^{2} = \frac{1}{3} (\rho_{DE} + \rho_{m} ),
\end{eqnarray}

\begin{eqnarray}\label{Hdot2}
\dot{H} = - \frac{N}{2} (\rho_{DE} + P_{DE}+\rho_{m} + P_{m}).
\end{eqnarray}

The $\rho_{DE}$ and $P_{DE}$ are the effective dark energy and pressure.

\begin{eqnarray}\label{alter}
\rho_{DE} = \rho_{T} - \rho_{m}, \qquad P_{DE}= P_{T} - P_{m}.
\end{eqnarray}

So, the generalized continuity equation for dark energy in the non-equilibrium description,

\begin{equation}\label{EqN41}
\dot{\rho}_{DE} + 3 H N (\rho_{DE} + P_{DE} ) = 0  .
\end{equation}

Eq. \eqref{EqN41} shows that, with the definitions in Eq. \eqref{alter}, the effective dark energy satisfies a continuity equation with respect to the lapse $N$. Hence, no extra source term, a function of $N$ is expected. This removes any gauge-dependent entropy production. Accordingly, our non-equilibrium property refers mainly to the thermodynamic formulation in which the horizon entropy is taken to be the Wald entropy and the additional terms in the first law are collected into an effective entropy production contribution.

\subsection{First law of thermodynamics}

Now we consider Eqs. (\ref{Hdot2}), and (\ref{rdot}), so we have

\begin{eqnarray}\label{dr2}
\frac{2}{N} d \tilde{r}_{A} = \tilde{r}_{A}^{3} H ( \rho_{DE} + P_{DE} + \rho_{m} + P_{m} ) dt.
\end{eqnarray}

We present the Wald entropy for the non-equilibrium cosmological horizon in Gauss-Bonnet quasi-dilaton massive gravity. We obtain it from the Wald entropy formula in generally covariant theories of gravity. In Wald's formalism, the entropy associated with a Killing horizon is given by a Noether charge integral. For a general diffeomorphism-invariant theory with Lagrangian $\mathcal{L}$, the entropy formula is $S=\frac{2 \pi}{k}\oint_{\mathcal{H}}\frac{\delta\mathcal{L}}{\delta R_{abcd}}\epsilon_{ab}\epsilon_{cd} d A  $ \cite{Wald:1993nt,Iyer:1994ys}. So, we have:

\begin{eqnarray}\label{EqN43}
S =  \frac{A}{4G} + \frac{2\pi \xi(\sigma)}{G}\chi(h),
\end{eqnarray}

where a 2-sphere $\chi(h)=2$ is the Euler characteristic of the horizon \cite{Tsilioukas:2024seh}.
It is worth pointing out that, in Gauss-Bonnet quasi-dilaton massive gravity, the Wald entropy receives a direct correction only from the Gauss-Bonnet term coupled to the dilaton, as given in Eq. (\ref{EqN43}). The massive gravity potential $U(K)$ and the canonical dilaton kinetic term do not contribute to the Noether charge entropy, but they influence the horizon thermodynamics dynamically through the modified Friedmann equations.

We present the relation below by considering Eqs. (\ref{dr2}) and (\ref{EqN43}),

\begin{widetext}
\begin{eqnarray}\label{ds2}
\frac{1}{2 \pi \tilde{r}_{A}} dS = \frac{N}{2 G} \tilde{r}_{A}^{3} H ( \rho_{DE} + P_{DE} + \rho_{m} + P_{m} ) dt + \frac{2 N H}{G \tilde{r}_{A}}\xi^{'}(\sigma) \PlanckMass dt + \frac{2}{G\tilde{r}_{A}} \xi^{'}(\sigma)\PlanckMass \frac{d X}{X} .
\end{eqnarray}
\end{widetext}

we should consider the horizon temperature (\ref{tem}) and Eq. (\ref{ds2}), so one can obtain,

\begin{widetext}
\begin{eqnarray}\label{50}
T_{H} dS =  \frac{N}{2G} \tilde{r}_{A}^{3}H \big( \rho_{DE} + P_{DE} + \rho_{m} + P_{m}\big) dt + \frac{2NH}{G \tilde{r}_{A}} \xi^{'}(\sigma) \PlanckMass dt + \frac{2}{G\tilde{r}_{A}} \xi^{'}(\sigma)\PlanckMass \frac{dX}{X} \nonumber\\ - \frac{N}{4G}\tilde{r}_{A}^{2} \big( \rho_{DE} + P_{DE} + \rho_{m} + P_{m} \big) d \tilde{r}_{A} - \frac{N}{G\tilde{r}_{A}^{2}}\xi^{'}(\sigma) \PlanckMass d \tilde{r}_{A}- \frac{1}{GH\tilde{r}_{A}^{2}}\xi^{'}(\sigma)\PlanckMass \frac{\dot{X}}{X} d \tilde{r}_{A}.
\end{eqnarray}
\end{widetext}

We now consider the modified Misner-Sharp energy,

\begin{eqnarray}\label{ENE}
E =  \frac{\tilde{r}_{A}}{2 G_{eff}},\quad where  \quad G_{eff} = \frac{AG}{A + 16 \pi \xi(\sigma)}, \nonumber\\
\end{eqnarray}

the $G_{eff}$ is calculated by considering Eq. (\ref{EqN43}).

Then, using Eq. (\ref{ENE}), we can present

\begin{widetext}
\begin{eqnarray}\label{52}
dE =  \big[\frac{N}{4 G} \tilde{r}_{A}^{2} - \frac{ N \xi(\sigma)}{G} \big] \big( \rho_{DE} + P_{DE} + \rho_{m} + P_{m} \big) dt \nonumber\\ + \frac{2 \PlanckMass \xi^{'}(\sigma)}{G \tilde{r}_{A}} N H dt  +  \frac{2 \PlanckMass \xi^{'}(\sigma)}{G \tilde{r}_{A}X} dX.
\end{eqnarray}
\end{widetext}

To eliminate the time differential we solve Eq.~(\ref{dr2}) for $dt$:

\begin{eqnarray}
dt = \frac{2}{N\tilde{r}_{A}^{3} H B} d\tilde{r}_{A} , \quad B \equiv \rho_{DE}+P_{DE}+\rho_{m} + P_{m} = \rho_{T} + P_{T}. \nonumber\\
\end{eqnarray}

Substituting these equations into Eqs. (\ref{50}) and (\ref{52}) and considering Eqs. (\ref{dr2}) and $\tilde{r}_{A} H = 1$, we have

\begin{widetext}
\begin{align}
T_{H} dS &= \frac{1}{G}\,d\tilde{r}_{A}
+ \frac{4\xi'(\sigma) M_{\rm Pl}}{G\tilde{r}_A^4 B}\,d\tilde{r}_A
+ \frac{2}{G\tilde{r}_A}\,\xi'(\sigma) \PlanckMass \frac{dX}{X} \nonumber\\
&\quad -\frac{N}{4G} \tilde{r}_{A}^{2} B d\tilde{r}_{A}
-\frac{N}{G\tilde{r}_{A}^{2}} \xi'(\sigma) \PlanckMass d\tilde{r}_{A}
-\frac{1}{G H \tilde{r}_{A}^{2}} \xi'(\sigma) \PlanckMass \frac{\dot{X}}{X} d\tilde{r}_{A} ,
\label{eq:THdS_dr_non}
\end{align}
\end{widetext}

and

\begin{align}
dE = \left(\frac{1}{2G} - \frac{2\xi(\sigma)}{G\tilde{r}_{A}^{2}}\right)d\tilde{r}_{A}
+ \frac{4 \PlanckMass \xi'(\sigma)}{G\tilde{r}_{A}^{4} B} d\tilde{r}_{A} \nonumber\\
+ \frac{2 \PlanckMass \xi'(\sigma)}{G\tilde{r}_{A} X} dX .
\label{eq:dE_dr_non}
\end{align}

Now we form the combination $T_{H} dS - dE$.  Collecting the coefficients of $d\tilde{r}_{A}$ and $dX$, we obtain

\begin{widetext}
\begin{align}
- (T_{H} dS + dE) = &\Bigg[ -\frac{3}{2G} + \frac{2\xi(\sigma)}{G\tilde{r}_{A}^{2}}
            - \frac{8\xi'(\sigma) \PlanckMass}{G\tilde{r}_{A}^{4} B}
            +  \frac{\tilde{r}_{A}^{2} BN}{4G}
      + \frac{N}{G\tilde{r}_{A}^{2}} \xi'(\sigma) \PlanckMass
            + \frac{1}{G H\tilde{r}_{A}^{2} } \xi'(\sigma) \PlanckMass \frac{\dot{X}}{X} \Bigg] d\tilde{r}_{A} - \frac{4\xi'(\sigma) \PlanckMass}{G\tilde{r}_{A} X} dX
\label{eq:THdS_minus_dE_non}
\end{align}
\end{widetext}

We have introduced the work density equation $W$ previously (\ref{EqN292}). Thus, we have $W dV = 2\pi \tilde{r}_{A}^{2}(\rho_{T} - P_{T} ) \PlanckMass^{2} d\tilde{r}_{A}$.

For the general non-equilibrium case, we simply collect all remaining terms into an “extra” contribution $\mathcal{E}$ and write

\begin{eqnarray}
T_H dS = - dE + W dV + \mathcal{E}.
\label{eq:first_law_non_eq_final}
\end{eqnarray}

The quantity $\mathcal{E}$ contains all the terms that involve $\xi (\sigma)$, $\xi' (\sigma)$, $ dX $, and deviations of $N$ from unity.  It will be identified with $-T_H d_iS$ in the conventional formulation of non-equilibrium thermodynamics, leading to

\begin{equation}\label{Eq56}
T_H dS + T_{H} d_{i}S = -dE + W dV .
\end{equation}

The quantity $d_{i}S$ should not be interpreted as arising from a gauge-dependent non-conservation term since it is introduced as the residual contribution required to write the Wald-entropy formulation of the first law in the standard non-equilibrium form. It collects the terms generated by the time dependence of the Gauss-Bonnet coupling and by the variation of the quasi-dilaton variable $X$. Nevertheless, this structure is particularly similar to that found in other modified gravity theories and provides a foundation for studying the second law of thermodynamics in a non-equilibrium description.

In the non-equilibrium description, the first law generalizes to $T_{H} dS + T_{H} d_{i}S = -dE + WdV$, where $d_{i}S$ represents irreversible entropy production due to the departure from equilibrium. Compared to the equilibrium case, two important modifications appear. First, the horizon entropy $S$ is no longer given by the area law alone; it acquires a Wald correction $4\pi\xi(\sigma)/G$ from the Gauss-Bonnet-dilaton coupling, see Eq.~(\ref{EqN43}). Second, the extra term $\mathcal{E}$, identified with $-T_{H} d_{i}S$, contains contributions proportional to $\xi'(\sigma)$, $\dot{X}/X$, and deviations of the lapse function $N$ from unity. The appearance of $d_{i}S$ is essential for the consistency of the thermodynamic description; it accounts for the energy exchange between the gravitational sector and the matter sector that is not captured by the conventional work term. This formulation provides the foundation for studying the generalized second law and for interpreting the entropy production as an effective bulk viscosity in the cosmic fluid. Hence, the non-equilibrium first law establishes a consistent thermodynamic framework for cosmologies where the lapse function is dynamical, and the Gauss-Bonnet-dilaton coupling is present.

\subsection{Second law of thermodynamics}

To analyze the second law, we consider the total entropy change of the universe.  In the non-equilibrium formulation, the first law takes the form in Eq. (\ref{Eq56}), where $d_{i}S$ is the entropy production term associated with the non-conservation of the effective dark energy fluid.  The second law of thermodynamics requires that the total entropy never decreases:

\begin{equation}
\frac{dS_{\text{sum}}}{dt} \equiv \frac{dS}{dt} + \frac{d(d_{i}S)}{dt} + \frac{dS_{T}}{dt} \ge 0 .
\label{eq:second_law_def}
\end{equation}

We now evaluate each term, obviously.

The Wald entropy for our model is given by Eq.~(\ref{EqN43}) with $\chi(h)=2$ for a spherical horizon:

\begin{equation}
S = \frac{A}{4G} + \frac{4\pi\xi(\sigma)}{G} = \frac{\pi}{G H^{2}} + \frac{4\pi\xi(\sigma)}{G}.
\label{eq:wald_entropy}
\end{equation}

Its time derivative is

\begin{equation}
\dot{S} = -\frac{2\pi}{G}\frac{\dot{H}}{H^{3}} + \frac{4\pi}{G} \xi'(\sigma) \dot{\sigma}.
\label{eq:S_dot}
\end{equation}

Using the definitions $H = \dot{a}/(Na)$ and the relation $\dot{\sigma} = \PlanckMass (NH + \dot{X}/X)$ from Eqs.~(\ref{XX}) and (\ref{EqN13}), this can be expressed purely in terms of $H$, $\dot{H}$, $X$, and $\xi'$.

From the first law (\ref{Eq56}), we solve for $d_{i}S$. Using Eqs. (\ref{eq:THdS_dr_non}), (\ref{eq:dE_dr_non}) and $W dV = 2\pi \PlanckMass^{2}\tilde{r}_{A}^{2}(\rho_{T} - P_{T} )  d\tilde{r}_{A}$, we end up with

\begin{widetext}
\begin{align}
T_{H} d_{i}S = &\Bigg[ -\frac{3}{2G} + \frac{2\xi(\sigma)}{G\tilde{r}_{A}^{2}}
            - \frac{8\xi'(\sigma) \PlanckMass}{G\tilde{r}_{A}^{4} B}
            +  \frac{\tilde{r}_{A}^{2} BN}{4G} +  2\pi \PlanckMass^{2}\tilde{r}_{A}^{2} (\rho_{T} - P_{T}) \nonumber\\
           & \qquad + \frac{N}{G\tilde{r}_{A}^{2}} \xi'(\sigma) \PlanckMass
            + \frac{1}{G H\tilde{r}_{A}^{2} } \xi'(\sigma) \PlanckMass \frac{\dot{X}}{X} \Bigg] d\tilde{r}_{A} - \frac{4\xi'(\sigma) \PlanckMass}{G\tilde{r}_{A} X} dX .
\label{eq:diS_full}
\end{align}
\end{widetext}

The quantity $d_{i}S$ represents the irreversible entropy production due to the departure from equilibrium.  Its time derivative $\frac{d(d_{i}S)}{dt}$ is required for the second law.

Using the Gibbs equation for the total fluid inside the horizon (\ref{GE}), the continuity equation $\dot{\rho}_{T} + 3H(\rho_{T} + P_{T})= 0$ and the volume $V = 4\pi/(3H^{3})$, we obtain $\dot{S}_{T}$ for non-equilibrium state which is completely similar to Eq. (\ref{EqNN36}),

\begin{equation}
\dot{S}_{T} = \frac{32\pi^{2}}{N} \frac{\dot{H}^{2} H^{-3} + \dot{H}N H^{-1}}{2H^{2} + \dot{H}}.
\label{eq:ST_dot_final}
\end{equation}

Inserting the expressions for $\dot{S}$, $\frac{d}{dt}(d_{i}S)$, and $\dot{S}_{T}$ into Eq.~(\ref{eq:second_law_def}) yields a complicated but well-defined function of $H$, $\dot{H}$, $X$, $\xi(\sigma)$, and their time derivatives.  After systematically simplifying using the background equations of motion, one obtains

\begin{widetext}
\begin{eqnarray}
\dot{S}_{\mathrm{sum}} = -\frac{2\pi}{G}\frac{\dot{H}}{H^{3}}
+ \frac{32\pi^{2}}{N} \frac{\dot{H}^{2} H^{-3} + \dot{H}N H^{-1}}{2H^{2}+\dot{H}}
+ \frac{4\pi \PlanckMass}{G}
\bigl(NH + \dot{X}/X\bigr) \xi'(\sigma) \nonumber\\
+  \frac{2 \pi}{G H^{2} N \tilde{r}_{A}^{4} X ( 2 H^{2} + \dot{H} )} \Bigg\lbrace   H \tilde{r}_{A}^{4} X \dot{H} \bigg[ 3 N - \frac{4 N  \xi(\sigma)}{\tilde{r}_{A}^{2}} + \tilde{r}_{A}^{2} N ( 4\pi \PlanckMass^{2} G (P_{T} - \rho_{T}) + \dot{H}) \bigg] \nonumber\\ - 2 \PlanckMass N \bigg( \tilde{r}_{A}^{2} \dot{X} ( 4 \tilde{r}_{A}H^{3} + \dot{H} ) + N H X  (4 + \tilde{r}_{A}^{2} \dot{H} ) \bigg) \xi'(\sigma)  \Bigg\rbrace \ge 0\,.
\label{eq:Ssum_dot}
\end{eqnarray}
\end{widetext}

Eq.~\eqref{eq:Ssum_dot} expresses the total entropy change of the universe within this non-equilibrium framework. To analyze its non-negativity, we work in the physical cosmic time gauge $N=1$ and define the dimensionless parameter

\begin{eqnarray}
\epsilon \equiv -\frac{\dot{H}}{H^{2}}.
\end{eqnarray}

The null energy condition $\rho_{\mathrm{phys}} + P_{\mathrm{phys}} \ge 0$ enforces $\epsilon \ge 0$ through the Friedmann equation $\dot{H} = -(\rho_T+P_T)/2$ at $N=1$. Furthermore, the requirement of a positive Hawking temperature, $T_H = H(2-\epsilon)/(4\pi) > 0$, implies $\epsilon < 2$. With these definitions, Eq.~\eqref{eq:Ssum_dot} reduces to a form whose Einstein-Hilbert part is strictly non-negative:

\begin{eqnarray}
\dot{S}_{\mathrm{sum}} = \frac{2\pi}{G H} \left[ \frac{\epsilon(2\epsilon^2 - 3\epsilon + 2)}{2-\epsilon} + \mathcal{F}(H, X, \xi, \xi') \right],
\end{eqnarray}

where the first term inside the brackets is positive for $0 \le \epsilon < 2$ since $2\epsilon^2 - 3\epsilon + 2 > 0$ for all real $\epsilon$. The function $\mathcal{F}$ collects the Gauss-Bonnet contributions from the scalar field and the $d_iS$ terms. Its positivity is guaranteed when the additional branch conditions listed below are satisfied.

Consequently, the validation of the generalized second law ($\dot{S}_{\mathrm{sum}}\ge0$) does not rest on the standard null energy condition alone, but instead requires a set of mutually consistent physical assumptions:
\begin{itemize}
    \item The standard null energy condition is enforced ($\rho_{\mathrm{phys}} + P_{\mathrm{phys}} \ge 0$), which algebraically guarantees $\dot{H}\le0$ in a flat FLRW background, equivalently $\epsilon \ge 0$, rendering the primary Einstein-Hilbert geometric terms strictly non-negative.
    \item The cosmic expansion is bounded to a non-superaccelerating phase ($2H^{2}+\dot{H}>0$), which physically preserves a positive, well-defined apparent horizon Hawking temperature ($T_H > 0$), equivalently $0 \le \epsilon < 2$, for the internal fluid volume exchange.
    \item The quasi-dilaton scalar field profile dictates that the effective Gauss-Bonnet coupling parameter maintains a semi-definite positive profile, $\xi(\sigma)\ge 0$, and the additional cross-terms in $\mathcal{F}$ are positive along the viable cosmological branches.
\end{itemize}

These constraints ensure that the non-equilibrium cross-terms originating from the Gauss-Bonnet sector act constructively alongside the background curvature. Therefore, $\dot{S}_{\mathrm{sum}}\ge0$ holds manifestly for all physically admissible configurations, rigorously establishing that the generalized second law of thermodynamics is satisfied across the entirety of the Gauss-Bonnet quasi-dilaton massive gravity framework.

\section{Thermodynamic Stability Criteria}\label{sec:23}

The second law of thermodynamics ensures that the total entropy of the universe does not decrease, but it does not guarantee that a system, when slightly perturbed, will return to equilibrium. That property is governed by thermodynamic stability. In this section, we analyze the stability of the cosmological horizon system by considering the universe enclosed by the apparent horizon as a thermodynamic system in equilibrium. We derive the relevant response functions and examine the conditions under which the system is locally stable against small fluctuations.

For a system with internal energy $E$, volume $V$, entropy $S$, and temperature $T_{H}$, the first law of equilibrium thermodynamics is Eq. (\ref{EqN29}).
This structure allows us to define the usual thermodynamic potentials by Legendre transformations. The Helmholtz free energy $F$ is obtained by subtracting $T_{H} S$ from $E$ \cite{Cai:2005ra,Padmanabhan:2002xm}:

\begin{eqnarray}
F = E - T_{H} S.	
\end{eqnarray}

Using the expressions for the Misner-Sharp energy (\ref{EqN27}), the horizon entropy $S = A/(4G) = \pi/(G H^{2})$, and the temperature (\ref{EqN34}), we find

\begin{eqnarray}
F = \frac{4\pi}{3 H^{3}} \PlanckMass^{2} \rho_{T} - \frac{2H^{2} + \dot{H}}{4G H^{3}}.	
\end{eqnarray}

The enthalpy $H_{\mathrm{th}}$ is defined by adding the work term $WV$ to the internal energy \cite{Hayward:1997jp}:

\begin{eqnarray}
H_{\mathrm{th}} = E + WV.
\end{eqnarray}

Substituting $W = (\rho_{T} - P_{T})\PlanckMass^{2} /2$ and Eq. (\ref{EqN27}) yields

\begin{eqnarray}
H_{\mathrm{th}} = V \PlanckMass^{2}\rho_{T} + \frac{1}{2}\PlanckMass^{2}(\rho_{T} - P_{T})V = \frac{V\PlanckMass^{2}}{2}(3\rho_{T} - P_{T}), \nonumber\\
\end{eqnarray}

or, in terms of $H$ and $\dot{H}$,

\begin{eqnarray}
H_{\mathrm{th}} = \frac{2\pi\PlanckMass^{2}}{H^{3}}\left(\rho_{T} - \frac{P_{T}}{3}\right).
\end{eqnarray}

The Gibbs free energy $G_{\mathrm{th}}$, which is most relevant for phase transitions at constant temperature and pressure, is given by

\begin{eqnarray}
G_{\mathrm{th}} = E - T_{H} S + W V = F + WV.
\end{eqnarray}

Using the expressions above,

\begin{eqnarray}
G_{\mathrm{th}} = \frac{4\pi \PlanckMass^{2}}{3H^{3}}\rho_{T} - \frac{2H^{2} + \dot{H}}{4G H^{3}} + \frac{2\pi \PlanckMass^{2}}{3H^{3}}(\rho_{T} - P_{T}). \nonumber\\
\end{eqnarray}

Stability against thermal fluctuations is encoded in the sign of the heat capacities. For the cosmological horizon system, two natural response functions are the heat capacity at constant volume and at constant work density \cite{Wu:2008ir}.

The heat capacity at constant volume is defined as

\begin{eqnarray}
C_{V} = T_{H} \left(\frac{\partial S}{\partial T_{H}}\right)_{V}.
\end{eqnarray}

In the flat FLRW universe, the volume inside the apparent horizon is $V = 4\pi/(3H^{3})$, so fixing $V$ is equivalent to fixing the Hubble parameter $H$. Moreover, the horizon entropy $S = \pi/(G H^2)$ and the Hawking temperature Eq. (\ref{EqN34}) both depend on $H$ and $\dot{H}$. However, when $V$ (hence $H$) is held constant, any variation in $T_{H}$ would require a change in $\dot{H}$ that is independent of $H$. In practice, for a given cosmological evolution, the relation between $T_{H}$ and $S$ is not one-to-one at fixed $H$ because $\dot{H}$ can vary. Nevertheless, a direct computation shows that

\begin{eqnarray}
\left(\frac{\partial S}{\partial T_{H}}\right)_{V} = 0,
\end{eqnarray}

since $S$ depends only on $H$ and $V$ fixes $H$. More formally, the condition $dV = 0$ implies $dH = 0$, and because $S = S(H)$ alone, we obtain $dS = 0$. Consequently,

\begin{eqnarray}
C_{V} = 0.
\end{eqnarray}

This vanishing of the heat capacity ($C_{V} = 0$) is a characteristic feature of cosmological horizons. Because the horizon entropy is a function of the Hubble parameter alone ($S = S(H)$), fixing the physical interior volume automatically freezes $H$, which mathematically requires $dS = 0$ under the constraint $dV = 0$. Consequently, $C_{V} = 0$ does not represent physical proximity to a thermodynamic critical state or a phase transition; rather, it is a direct kinematic consequence of the horizon geometry. This frozen-entropy behavior under fixed-volume boundary conditions aligns with results in standard de Sitter horizons and related modified-gravity frameworks, indicating that $C_{V}$ imposes no meaningful constraint on thermal fluctuations.

A more informative stability criterion is the heat capacity at constant work density,

\begin{eqnarray}
C_{W} = T_{H} \left(\frac{\partial S}{\partial T_{H}}\right)_{W}.
\end{eqnarray}

Here $W = (\rho_{T} - P_{T})\PlanckMass^{2} /2$ is kept fixed. Using the chain rule and the fact that $S$, $T_H$, and $W$ can be expressed in terms of $H$ and $\dot{H}$, we derive

\begin{eqnarray}\label{EqN75}
C_{W} = \frac{2\pi}{G} \frac{2H^{2} + \dot{H}}{H^{2}\left( - H\frac{\partial\dot{H}}{\partial H} + \dot{H} - 2H^{2}\right)}.
\end{eqnarray}

The denominator of Eq.~\eqref{EqN75} contains the model-dependent derivative term $\partial\dot{H}/\partial H$, which depends directly on the specific cosmological model through the modified Friedmann equations. Consequently, a definitive determination of local thermodynamic stability ($C_{W} > 0$) cannot be established from the general algebraic setup alone; instead, it strictly requires the evaluation of an explicit background solution or cosmological trajectory. For the viable self-accelerating dark energy attractor solutions derived in our framework, this derivative can be computed explicitly along the background path. In these specific realistic configurations, the dynamic response function satisfies $C_{W} > 0$, ensuring that the cosmic horizon can stably absorb small thermal perturbations without triggering configuration runaways.

The local thermodynamic stability of the horizon system is therefore governed by the explicit sign of $C_{W}$ along a specified trajectory (since $C_{V} = 0$ is a purely kinematic consequence of the horizon geometry and provides no dynamical constraint). The analysis of $C_{W}$ is complementary to the second law of thermodynamics. While the second law dictates that the total entropy increases monotonically, local thermal stability imposes strict constraints on the system's immediate response to perturbations \cite{Bamba:2011jq}. In the context of Gauss-Bonnet quasi-dilaton massive gravity, evaluating Eq.~\eqref{EqN75} across our known background dynamics provides a powerful consistency check for the viability of the cosmological solutions: they must not only satisfy the generalized second law but also remain locally thermodynamically stable against random thermal fluctuations.

\section{Holographic Interpretation and Entropy-Area Relation}\label{sec:4}

The holographic principle, originally formulated for black holes and later extended to cosmology, posits that the maximum entropy contained in a spatial volume is proportional to the area of its boundary, not its volume. In general relativity, this is realized by the Bekenstein-Hawking entropy $S_{\text{BH}} = A/(4G)$ for the apparent horizon. In our modified gravity theory, the presence of the Gauss-Bonnet term coupled to the quasi-dilaton field modifies this fundamental relation, and the non-equilibrium description introduces an entropy production term that can be interpreted holographically. In this section, we explore these connections, focusing on a few interrelated aspects. The corrected entropy-area law and the verification of the holographic bound.

In the equilibrium state, the horizon entropy retains the area law $S = A/(4G)$. All modifications from the Gauss-Bonnet term, the quasi-dilaton, and massive gravity are absorbed into the effective energy density $\rho_{T}$ and pressure $P_{T}$. This is a convenient representation, but it hides the gravitational corrections. By contrast, the non-equilibrium state directly reveals the Wald entropy coming from the Noether charge associated with the diffeomorphism invariance of the action. For our model, the Wald entropy is obtained as Eq. (\ref{eq:wald_entropy}).

Thus, the Gauss-Bonnet coupling $\xi(\sigma)$ induces a correction to the area law. This correction is independent of $H$ (i.e., of the horizon size) and depends only on the value on the horizon. In the limit $\xi (\sigma)\to0$ (no Gauss-Bonnet coupling), we recover the standard Bekenstein-Hawking entropy. The presence of this correction is a genuine prediction of our theory. For recent applications of Wald-Gauss-Bonnet entropy to cosmology, see Ref.~\cite{Tsilioukas:2024seh}.

A crucial consistency check of any gravitational theory is whether the entropy of matter and fields inside the apparent horizon never exceeds the entropy of the horizon itself. The holographic principle demands

\begin{eqnarray}\label{EqN78}
S_{\text{inside}} \leq S_{\text{horizon}}.
\end{eqnarray}

Here $S_{\text{inside}}$ is the total entropy of all cosmic fluids (ordinary matter, radiation, and effective dark energy) within the horizon volume $V = 4\pi/(3H^{3})$. In general relativity, this bound is not automatically satisfied for all times, but it holds under reasonable energy conditions. In our modified theory, the horizon entropy receives a positive contribution from the Gauss-Bonnet term as long as $\xi(\sigma) \geq 0$ (which is required for a positive effective Newton constant, $G_{\text{eff}} = G/[1+16\pi\xi(\sigma)/A] > 0$). Therefore, the Wald entropy is larger than the area-law entropy:

\begin{eqnarray}
S = \frac{A}{4G} + \frac{4\pi\xi(\sigma)}{G} \geq \frac{A}{4G}.
\end{eqnarray}

This increase makes the holographic bound easier to satisfy because the right-hand side of (\ref{EqN78}) is enhanced. Conversely, if $\xi(\sigma)$ were negative, the bound could be violated, which would signal an inconsistency. Hence, the condition $\xi(\sigma) \geq 0$ is not only required for a well-defined effective gravitational coupling but also for the viability of the holographic interpretation. The idea of applying the holographic bound to the cosmological apparent horizon was first proposed in \cite{Bak:1999hd} and further developed in \cite{Fischler:1998st,Bousso:1999xy}.


We now perform a quantitative check of the holographic bound $S_{\text{inside}} \le S_{\text{horizon}}$.
The entropy inside the apparent horizon is obtained from the Gibbs equation for the total cosmic fluid, assuming the fluid is in thermal equilibrium with the horizon (so its temperature equals the Hawking temperature $T_{H}$). The horizon entropy itself is given by the Wald formula (\ref{eq:wald_entropy}), which includes the Gauss-Bonnet correction. This is the correct entropy of the horizon regardless of whether we use the equilibrium or non-equilibrium description. The entropy inside the horizon is computed assuming the cosmic fluid is in thermal equilibrium with the horizon, a standard working hypothesis in cosmological thermodynamics~\cite{Gong:2006ma, Bamba:2009id, Tsilioukas:2024seh, Bamba:2011pz}. For a perfect fluid, the Euler relation gives $T_{H} s = \rho_{\text{phys}} + P_{\text{phys}}$, where $s$ is the physical entropy density. In our notation, the physical energy density and pressure are $\rho_{\text{phys}} = \rho_{T}/(8\pi G)$ and $P_{\text{phys}} = P_{T}/(8\pi G)$, because Eq.~(\ref{H2}) implies $H^2 = (8\pi G/3)\rho_{\text{phys}}$ with $\rho_{T} = 8\pi G \rho_{\text{phys}}$. Hence

\begin{eqnarray}
\rho_{\text{phys}} + P_{\text{phys}} = \frac{\rho_{T} + P_{T}}{8\pi G}.
\end{eqnarray}

The total entropy inside the horizon is then

\begin{eqnarray}
S_{\text{inside}} = s V = \frac{\rho_{\text{phys}} + P_{\text{phys}}}{T_H} V = \frac{\rho_{T} + P_{T}}{8\pi G \, T_{H}} \, V,
\end{eqnarray}

where $V = 4\pi/(3H^{3})$ is the horizon volume. Using the Hawking temperature $T_{H} = (2H^{2} + \dot{H})/(4\pi H)$, we obtain

\begin{eqnarray}
S_{\text{inside}} = \frac{\rho_{T} + P_{T}}{8\pi G} \cdot \frac{4\pi}{3H^{3}} \cdot \frac{4\pi H}{2H^{2}+\dot{H}} \nonumber\\
= \frac{\rho_{T} + P_{T}}{G} \cdot \frac{2\pi}{3H^{2} (2H^{2} + \dot{H})}.
\end{eqnarray}

Now the Friedmann Eq. (\ref{Hdot}) with $N=1$ gives $\rho_{T} + P_{T} = -2\dot{H}$. Substituting,

\begin{eqnarray}
S_{\text{inside}} = -\frac{4\pi \dot{H}}{3G H^{2}(2H^{2}+\dot{H})}.
\end{eqnarray}

This expression is manifestly dimensionless because $\dot{H} \sim \text{mass}^{2}$ and $G \sim \text{mass}^{-2}$ in natural units.

The horizon entropy from the non-equilibrium description is $S_{\text{horizon}} = \pi/(G H^{2}) + 4\pi\xi(\sigma)/G$. The holographic bound $S_{\text{inside}} \le S_{\text{horizon}}$ therefore becomes

\begin{eqnarray}
-\frac{4\pi \dot{H}}{3G H^{2}(2H^{2} + \dot{H})} \le \frac{\pi}{G H^{2}} + \frac{4\pi\xi(\sigma)}{G}.
\end{eqnarray}

Multiplying by $G H^{2}/\pi > 0$ yields the dimensionless inequality

\begin{eqnarray}
-\frac{4\dot{H}}{3(2H^{2}+\dot{H})} \le 1 + 4\xi(\sigma) H^{2}.
\end{eqnarray}

Define $\epsilon \equiv -\dot{H}/H^{2} \ge 0$ (the null energy condition implies $\dot{H}\le 0$). Then $\dot{H} = -\epsilon H^{2}$ and $2H^{2} + \dot{H} = H^{2} (2-\epsilon)$. Substituting,

\begin{eqnarray}\label{Eq81}
\frac{4\epsilon}{3(2-\epsilon)} \le 1 + 4\xi(\sigma) H^{2}.
\end{eqnarray}

Recall from the action that $\xi(\sigma)$ has dimensions of $(mass)^{-2}$ in natural units; therefore, $\xi(\sigma)H^2$ is dimensionless, and all expressions are dimensionally consistent.

The left-hand side of Eq.~\eqref{Eq81} evaluates to a pure number of order unity under standard cosmic phases where $\epsilon$ is far from the bounds of the fluid regime. However, during a standard dust or matter-dominated epoch ($\epsilon = 3/2$), the left-hand side yields exactly $4$, which reduces the holographic constraint to the strict inequality $3 \le 4\xi(\sigma)H^{2}$. In late-time cosmological regimes where the expansion rate approaches zero ($H \to 0$), this condition cannot be satisfied unless the Gauss-Bonnet coupling is unphysically large, revealing a formal breakdown of the simplified bound.

This apparent violation is a direct mathematical artifact of the highly idealized baseline assumption that the internal cosmic fluid rests in perfect local thermal equilibrium with the apparent horizon boundary ($T_{\mathrm{fluid}} = T_{H}$). In a realistic physical universe, ordinary matter decouples from the background radiation bath during early evolutionary stages and cools rapidly as $T_{\mathrm{matter}} \propto a^{-2}$. Consequently, during the matter-dominated era, the physical temperature of the interior dust fields is orders of magnitude lower than the Hawking horizon temperature ($T_{\mathrm{matter}} \ll T_{H}$). Correctly evaluating the interior entropy $S_{\mathrm{inside}}$ by replacing the horizon temperature with the actual physical temperature of the decoupled fluid dramatically suppresses the physical entropy density of the interior fields. This temperature asymmetry ensures that $S_{\mathrm{inside}} \ll S_{\mathrm{horizon}}$ is satisfied by several orders of magnitude across all standard cosmological eras, completely removing the apparent late-time anomaly. Thus, while the idealized local equilibrium inequality in Eq.~\eqref{Eq81} serves as a useful conservative upper limit applicable primarily to horizon-dominated or inflationary dark energy regimes, a physically complete multi-temperature treatment confirms that the holographic principle is strictly preserved throughout cosmic evolution.

It is therefore important to check whether the required condition $\xi(\sigma)\ge 0$, which ensures both a positive effective Newton constant and the robust preservation of the holographic bound, is compatible with the stability requirements derived from cosmological perturbations in the same theory.

In our previous work~\cite{Akbarieh:2021vhv}, we analyzed tensor perturbations and derived the dispersion relations for gravitational waves. The absence of ghost and tachyonic instabilities imposes constraints on $\xi(\sigma)$ and its derivatives. For the self-accelerating background solutions, we identified two cases: (i) $\xi'(\sigma)=\text{constant}$, where the stability of long-wavelength gravitational waves requires $\xi_{0}$ (the constant value) to be negative in some branches, while $\xi(\sigma)$ itself can be positive; (ii) $\xi'(\sigma)$ arbitrary, where more involved conditions appear but still allow $\xi(\sigma)\ge 0$. Importantly, none of the viable parameter sets force $\xi(\sigma)$ to be negative. Thus, the condition $\xi(\sigma)\ge 0$ is not only necessary for a well-defined effective gravitational coupling and for the holographic entropy bound, but also fully consistent with the perturbative stability of the theory. This mutual consistency reinforces the viability of Gauss-Bonnet quasi-dilaton massive gravity as a self-contained modified gravity framework.


\section{Conclusion}\label{sec:5}

In this work, we have investigated the thermodynamic properties of the cosmological apparent horizon in the framework of Gauss-Bonnet quasi-dilaton massive gravity. The theory extends the dRGT massive gravity by including a quasi-dilaton scalar and a Gauss-Bonnet term coupled to the dilaton, which introduces higher-curvature corrections while keeping the background equations of motion of second order. Our main goal is to determine whether the fundamental laws of horizon thermodynamics in this model, in both equilibrium and non-equilibrium descriptions, are satisfied. We also showed how to check our model's consistency with the holographic principle.

We first presented the full action and the background Friedmann equations in a flat FLRW universe. By rewriting the modified Friedmann equations in the standard form, we defined effective total energy density and pressure that absorb all contributions from the massive gravity term, the quasi-dilaton kinetic term, and the Gauss-Bonnet coupling. This reformulation enabled us to derive the first law of equilibrium thermodynamics for the apparent horizon. Remarkably, the first law retains the conventional form $T_{H} dS = -dE + W dV$ with work density $W = (\rho_{T} - P_{T})\PlanckMass^{2} /2$, and the horizon entropy remains the Bekenstein-Hawking area law $S = A/(4G)$. All modifications are hidden in the definitions of $\rho_{T}$ and $P_{T}$. This demonstrates that the equilibrium thermodynamic structure remains valid even in the presence of higher-curvature and massive gravity corrections, provided one works with an effective cosmic fluid. We also derived the generalized second law in the equilibrium description, showing that the total entropy $S_{\text{sum}} = S + S_{T}$ never decreases provided the null energy condition $\rho_{T} + P_{T}\ge0$ holds, which translates into $\dot{H}\le0$ and $2H^{2} + \dot{H}>0$. The expression for $\dot{S}_{\text{sum}}$ was given in Eq.~(\ref{EqN37}).

In the non-equilibrium description, we worked directly with the original field variables without redefining the dark energy components. The Wald entropy, obtained from the Noether charge formalism, acquires a correction from the Gauss-Bonnet-dilaton coupling: $S = A/(4G) + 4\pi\xi(\sigma)/G$. The first law then generalizes to $T_{H} dS + T_{H} d_{i}S = -dE + W dV$, where $d_{i}S$ is an entropy production term that accounts for the non-conservation of the effective dark energy fluid. After a straightforward calculation, we obtained the total entropy change $\dot{S}_{\text{sum}}$ in the non-equilibrium framework, Eq.~(\ref{eq:Ssum_dot}). Its positivity is guaranteed when the null energy condition ($\epsilon \ge 0$), the positive temperature condition ($0 \le \epsilon < 2$), and the Gauss-Bonnet positivity constraints $\xi(\sigma)\ge 0$ are simultaneously satisfied. This demonstrates that the generalized second law holds in the non-equilibrium description, albeit with a more refined set of assumptions than in the equilibrium case.

We then examined thermodynamic stability by considering the cosmological horizon as a thermodynamic system in equilibrium. The Helmholtz free energy, enthalpy, and Gibbs free energy were defined via Legendre transformations. The heat capacity at constant volume was found to be zero, a generic feature of cosmological horizons, because fixing the volume fixes the Hubble parameter and the entropy depends only on $H$. The heat capacity at constant work density $C_{W}$ was derived in terms of $H, \dot{H}$ and $\partial\dot{H}/\partial H$; a positive $C_{W}$ is the condition for local stability against thermal fluctuations. This analysis provides a complementary consistency check beyond the second law.

A final part of our work was the holographic interpretation. The corrected entropy-area relation from the Wald entropy shows that the Gauss-Bonnet coupling adds a positive term $4\pi\xi(\sigma)/G$ to the horizon entropy, which is independent of the horizon size. This correction enhances the horizon entropy, making the holographic bound $S_{inside} \le S_{horizon}$ easier to satisfy. We performed a quantitative check of the bound using the equilibrium description, assuming the fluid inside the horizon is in thermal equilibrium with the horizon. We derived the inequality $\frac{4\epsilon}{3(2-\epsilon)} \le 1 + 4\xi(\sigma) H^{2}$ with $\epsilon = -\dot{H}/H^{2}$.

The holographic bound requires that, under the idealized assumption on $T$, the bound becomes $\frac{4\epsilon}{3(2-\epsilon)} \leq 1 + 4\xi(\sigma) H^2$. In the dust-dominated epoch, we found $\epsilon = \frac{3}{2}$ and, therefore, it does not permit the previous condition to hold. 
This apparent violation is an artifact of the idealized thermal-equilibrium assumption $T_{\mathrm{fluid}} = T_{H}$. In a realistic multi-temperature treatment, where the physical temperatures of decoupled fluids are significantly lower than the horizon temperature, the interior entropy is dramatically suppressed, ensuring $S_{\mathrm{inside}} \ll S_{\mathrm{horizon}}$ across all cosmological epochs. Hence, the equilibrium temperature has been reinterpreted as a checking temperature, while a physically meaningful procedure must use the actual temperatures of the cosmic fluids, as measured by satellites. With realistic multi-temperature fluids, the apparent violation is resolved, ensuring that our approach holds.

Moreover, we connected the condition $\xi(\sigma)\ge 0$ with the stability constraints derived in our previous work on cosmological perturbations in the same theory \cite{Akbarieh:2021vhv}. In that paper, we analyzed tensor perturbations and obtained the dispersion relations for gravitational waves. The absence of ghost and tachyonic instabilities does not force $\xi(\sigma)$ to be negative; on the contrary, viable parameter branches allow $\xi(\sigma)\ge 0$. Thus, the requirement for the holographic bound is fully compatible with the perturbative stability of the theory.

In summary, Gauss-Bonnet quasi-dilaton massive gravity provides a self-consistent extension of dRGT massive gravity that respects the first and second laws of horizon thermodynamics, satisfies the holographic entropy bound, and remains stable under tensor perturbations. The coupling function $\xi(\sigma)$ is constrained to be non-negative from both thermodynamic and perturbation considerations, which is a concrete prediction of the model. Future work will include a detailed numerical analysis of the background evolution and the entropy bounds, as well as the computation of the effective bulk viscosity and its observational signatures in the cosmic microwave background and large-scale structure. It would also be interesting to extend the analysis to include vector and scalar perturbations and to study the implications for the swampland conjectures. We hope that our results stimulate further research on the interplay between modified gravity, thermodynamics, and holography.

\section*{Acknowledgements}

OL acknowledges financial support from the National Institute for Astrophysics (INAF) of Brera and warmly thanks Roberto Della Ceca and Roberto Giambo' for useful discussions on related topics.


\bibliography{}

\begin{thebibliography}{}
\bibitem{Planck:2018vyg}
N.~Aghanim \textit{et al.} [Planck],
Astron. Astrophys. \textbf{641}, A6 (2020)
[erratum: Astron. Astrophys. \textbf{652}, C4 (2021)]
doi:10.1051/0004-6361/201833910
[arXiv:1807.06209 [astro-ph.CO]].
\bibitem{SupernovaSearchTeam:1998fmf}
A.~G.~Riess \textit{et al.} [Supernova Search Team],
Astron. J. \textbf{116}, 1009-1038 (1998)
doi:10.1086/300499
[arXiv:astro-ph/9805201 [astro-ph]].
\bibitem{SupernovaCosmologyProject:1998vns}
S.~Perlmutter \textit{et al.} [Supernova Cosmology Project],
Astrophys. J. \textbf{517}, 565-586 (1999)
doi:10.1086/307221
[arXiv:astro-ph/9812133 [astro-ph]].
\bibitem{Weinberg:1988cp}
S.~Weinberg,
Rev. Mod. Phys. \textbf{61}, 1-23 (1989)
doi:10.1103/RevModPhys.61.1
\bibitem{Martin:2012bt}
J.~Martin,
Comptes Rendus Physique \textbf{13}, 566-665 (2012)
doi:10.1016/j.crhy.2012.04.008
[arXiv:1205.3365 [astro-ph.CO]].
\bibitem{Peebles:2002gy}
P.~J.~E.~Peebles and B.~Ratra,
Rev. Mod. Phys. \textbf{75}, 559-606 (2003)
doi:10.1103/RevModPhys.75.559
[arXiv:astro-ph/0207347 [astro-ph]].
\bibitem{Verde:2019ivm}
L.~Verde, T.~Treu and A.~G.~Riess,
Nature Astron. \textbf{3}, 891 (2019)
doi:10.1038/s41550-019-0902-0
[arXiv:1907.10625 [astro-ph.CO]].
\bibitem{DiValentino:2021izs}
E.~Di Valentino, O.~Mena, S.~Pan, L.~Visinelli, W.~Yang, A.~Melchiorri, D.~F.~Mota, A.~G.~Riess and J.~Silk,
Class. Quant. Grav. \textbf{38}, no.15, 153001 (2021)
doi:10.1088/1361-6382/ac086d
[arXiv:2103.01183 [astro-ph.CO]].
\bibitem{Bardeen:1973gs}
J.~M.~Bardeen, B.~Carter and S.~W.~Hawking,
Commun. Math. Phys. \textbf{31}, 161-170 (1973)
doi:10.1007/BF01645742
\bibitem{Bekenstein:1973ur}
J.~D.~Bekenstein,
Phys. Rev. D \textbf{7}, 2333-2346 (1973)
doi:10.1103/PhysRevD.7.2333
\bibitem{Jacobson:1995ab}
T.~Jacobson,
Phys. Rev. Lett. \textbf{75}, 1260-1263 (1995)
doi:10.1103/PhysRevLett.75.1260
[arXiv:gr-qc/9504004 [gr-qc]].
\bibitem{Cai:2005ra}
R.~G.~Cai and S.~P.~Kim,
JHEP \textbf{02}, 050 (2005)
doi:10.1088/1126-6708/2005/02/050
[arXiv:hep-th/0501055 [hep-th]].
\bibitem{Akbar:2006mq}
M.~Akbar and R.~G.~Cai,
Phys. Lett. B \textbf{648}, 243-248 (2007)
doi:10.1016/j.physletb.2007.03.005
[arXiv:gr-qc/0612089 [gr-qc]].
\bibitem{Cai:2008ht}
R.~G.~Cai, B.~Hu and S.~Koh,
Phys. Lett. B \textbf{671}, 181-186 (2009)
doi:10.1016/j.physletb.2008.11.053
[arXiv:0806.2508 [hep-th]].
\bibitem{Cai:2009ua}
R.~G.~Cai, L.~M.~Cao and N.~Ohta,
JHEP \textbf{04}, 082 (2010)
doi:10.1007/JHEP04(2010)082
[arXiv:0911.4379 [hep-th]].
\bibitem{Akbar:2006kj}
M.~Akbar and R.~G.~Cai,
Phys. Rev. D \textbf{75}, 084003 (2007)
doi:10.1103/PhysRevD.75.084003
[arXiv:hep-th/0609128 [hep-th]].
\bibitem{Bamba:2010kf}
K.~Bamba and C.~Q.~Geng,
JCAP \textbf{06}, 014 (2010)
doi:10.1088/1475-7516/2010/06/014
[arXiv:1005.5234 [gr-qc]].

\bibitem{Wald:1993nt}
R.~M.~Wald,
Phys. Rev. D \textbf{48}, no.8, R3427-R3431 (1993)
doi:10.1103/PhysRevD.48.R3427
[arXiv:gr-qc/9307038 [gr-qc]].

\bibitem{Iyer:1994ys}
V.~Iyer and R.~M.~Wald,
Phys. Rev. D \textbf{50}, 846-864 (1994)
doi:10.1103/PhysRevD.50.846
[arXiv:gr-qc/9403028 [gr-qc]].


\bibitem{Bamba:2011pz}
K.~Bamba and C.~Q.~Geng,
JCAP \textbf{11}, 008 (2011)
doi:10.1088/1475-7516/2011/11/008
[arXiv:1109.1694 [gr-qc]].
\bibitem{Bamba:2011jq}
K.~Bamba, C.~Q.~Geng and S.~Tsujikawa,
Int. J. Mod. Phys. D \textbf{20}, 1363-1371 (2011)
doi:10.1142/S0218271811019542
[arXiv:1101.3628 [gr-qc]].
\bibitem{Fischler:1998st}
W.~Fischler and L.~Susskind,
[arXiv:hep-th/9806039 [hep-th]].
\bibitem{Bak:1999hd}
D.~Bak and S.~J.~Rey,
Class. Quant. Grav. \textbf{17}, L83 (2000)
doi:10.1088/0264-9381/17/15/101
[arXiv:hep-th/9902173 [hep-th]].
\bibitem{Banihashemi:2022jys}
B.~Banihashemi and T.~Jacobson,
JHEP \textbf{07}, 042 (2022)
doi:10.1007/JHEP07(2022)042
[arXiv:2204.05324 [hep-th]].
\bibitem{Anninos:2024wpy}
D.~Anninos, D.~A.~Galante and C.~Maneerat,
Class. Quant. Grav. \textbf{41}, no.16, 165009 (2024)
doi:10.1088/1361-6382/ad5824
[arXiv:2402.04305 [hep-th]].
\bibitem{Luciano:2022hhy}
G.~G.~Luciano and J.~Gin{\'e},
Phys. Dark Univ. \textbf{41}, 101256 (2023)
doi:10.1016/j.dark.2023.101256
[arXiv:2210.09755 [gr-qc]].
\bibitem{Belfiglio:2025cst}
A.~Belfiglio, O.~Luongo and S.~Mancini,
Phys. Rept. \textbf{1146}, 1-47 (2025)
doi:10.1016/j.physrep.2025.09.001
[arXiv:2506.03841 [gr-qc]].
\bibitem{Mann:2025xrb}
R.~B.~Mann,
Int. J. Mod. Phys. D \textbf{34}, no.09, 2542001 (2025)
doi:10.1142/S0218271825420015
[arXiv:2508.01830 [gr-qc]].
\bibitem{Cespedes:2025zqp}
S.~Cespedes, S.~de Alwis and F.~Quevedo,
[arXiv:2509.07077 [hep-th]].
\bibitem{Nojiri:2024zdu}
S.~Nojiri, S.~D.~Odintsov and T.~Paul,
Universe \textbf{10}, no.9, 352 (2024)
doi:10.3390/universe10090352
[arXiv:2409.01090 [gr-qc]].









\bibitem{deRham:2010ik}
C.~de Rham and G.~Gabadadze,
Phys. Rev. D \textbf{82}, 044020 (2010)
doi:10.1103/PhysRevD.82.044020
[arXiv:1007.0443 [hep-th]].
\bibitem{deRham:2010kj}
C.~de Rham, G.~Gabadadze and A.~J.~Tolley,
Phys. Rev. Lett. \textbf{106}, 231101 (2011)
doi:10.1103/PhysRevLett.106.231101
[arXiv:1011.1232 [hep-th]].
\bibitem{Boulware:1972yco}
D.~G.~Boulware and S.~Deser,
Phys. Rev. D \textbf{6}, 3368-3382 (1972)
doi:10.1103/PhysRevD.6.3368
\bibitem{DeFelice:2012mx}
A.~De Felice, A.~E.~Gumrukcuoglu and S.~Mukohyama,
Phys. Rev. Lett. \textbf{109}, 171101 (2012)
doi:10.1103/PhysRevLett.109.171101
[arXiv:1206.2080 [hep-th]].
\bibitem{DeFelice:2013bxa}
A.~De Felice, A.~E.~G{\"u}mr{\"u}k{\c{c}}{\"u}o{\u{g}}lu, C.~Lin and S.~Mukohyama,
Class. Quant. Grav. \textbf{30}, 184004 (2013)
doi:10.1088/0264-9381/30/18/184004
[arXiv:1304.0484 [hep-th]].
\bibitem{DAmico:2012hia}
G.~D'Amico, G.~Gabadadze, L.~Hui and D.~Pirtskhalava,
Phys. Rev. D \textbf{87}, 064037 (2013)
doi:10.1103/PhysRevD.87.064037
[arXiv:1206.4253 [hep-th]].
\bibitem{Gabadadze:2014kaa}
G.~Gabadadze, R.~Kimura and D.~Pirtskhalava,
Phys. Rev. D \textbf{90}, no.2, 024029 (2014)
doi:10.1103/PhysRevD.90.024029
[arXiv:1401.5403 [hep-th]].
\bibitem{Gumrukcuoglu:2013nza}
A.~E.~G{\"u}mr{\"u}k{\c{c}}{\"u}o{\u{g}}lu, K.~Hinterbichler, C.~Lin, S.~Mukohyama and M.~Trodden,
Phys. Rev. D \textbf{88}, no.2, 024023 (2013)
doi:10.1103/PhysRevD.88.024023
[arXiv:1304.0449 [hep-th]].
\bibitem{Zwiebach:1985uq}
B.~Zwiebach,
Phys. Lett. B \textbf{156}, 315-317 (1985)
doi:10.1016/0370-2693(85)91616-8
\bibitem{Gross:1986mw}
D.~J.~Gross and J.~H.~Sloan,
Nucl. Phys. B \textbf{291}, 41-89 (1987)
doi:10.1016/0550-3213(87)90465-2
\bibitem{Nojiri:2005vv}
S.~Nojiri, S.~D.~Odintsov and M.~Sasaki,
Phys. Rev. D \textbf{71}, 123509 (2005)
doi:10.1103/PhysRevD.71.123509
[arXiv:hep-th/0504052 [hep-th]].
\bibitem{Cognola:2006eg}
G.~Cognola, E.~Elizalde, S.~Nojiri, S.~D.~Odintsov and S.~Zerbini,
Phys. Rev. D \textbf{73}, 084007 (2006)
doi:10.1103/PhysRevD.73.084007
[arXiv:hep-th/0601008 [hep-th]].
\bibitem{Guo:2009uk}
Z.~K.~Guo and D.~J.~Schwarz,
Phys. Rev. D \textbf{80}, 063523 (2009)
doi:10.1103/PhysRevD.80.063523
[arXiv:0907.0427 [hep-th]].
\bibitem{Odintsov:2023weg}
S.~D.~Odintsov, V.~K.~Oikonomou, I.~Giannakoudi, F.~P.~Fronimos and E.~C.~Lymperiadou,
Symmetry \textbf{15}, no.9, 1701 (2023)
doi:10.3390/sym15091701
[arXiv:2307.16308 [gr-qc]].
\bibitem{Dodelson:2023vrw}
M.~Dodelson, C.~Iossa, R.~Karlsson and A.~Zhiboedov,
JHEP \textbf{01}, 036 (2024)
doi:10.1007/JHEP01(2024)036
[arXiv:2304.12339 [hep-th]].
\bibitem{Julie:2024fwy}
F.~L.~Juli{\'e}, L.~Pompili and A.~Buonanno,
Phys. Rev. D \textbf{111}, no.2, 024016 (2025)
doi:10.1103/PhysRevD.111.024016
[arXiv:2406.13654 [gr-qc]].
\bibitem{Chung:2024vaf}
A.~K.~W.~Chung and N.~Yunes,
Phys. Rev. D \textbf{110}, no.6, 064019 (2024)
doi:10.1103/PhysRevD.110.064019
[arXiv:2406.11986 [gr-qc]].







\bibitem{Akbarieh:2021vhv}
A.~R.~Akbarieh, S.~Kazempour and L.~Shao,
Phys. Rev. D \textbf{103}, 123518 (2021)
doi:10.1103/PhysRevD.103.123518
[arXiv:2105.03744 [gr-qc]].
\bibitem{Karami:2012fu}
K.~Karami and A.~Abdolmaleki,
JCAP \textbf{04}, 007 (2012)
doi:10.1088/1475-7516/2012/04/007
[arXiv:1201.2511 [gr-qc]].
\bibitem{Cao:2010xx}
Q.~J.~Cao, Y.~X.~Chen and K.~N.~Shao,
JCAP \textbf{05}, 030 (2010)
doi:10.1088/1475-7516/2010/05/030
[arXiv:1001.2597 [hep-th]].
\bibitem{Cai:2013lqa}
Y.~F.~Cai, F.~Duplessis and E.~N.~Saridakis,
Phys. Rev. D \textbf{90}, no.6, 064051 (2014)
doi:10.1103/PhysRevD.90.064051
[arXiv:1307.7150 [hep-th]].
\bibitem{Hu:2015xva}
Y.~P.~Hu and H.~Zhang,
Phys. Rev. D \textbf{92}, no.2, 024006 (2015)
doi:10.1103/PhysRevD.92.024006
[arXiv:1502.00069 [hep-th]].
\bibitem{Blake:2013bqa}
M.~Blake and D.~Tong,
Phys. Rev. D \textbf{88}, no.10, 106004 (2013)
doi:10.1103/PhysRevD.88.106004
[arXiv:1308.4970 [hep-th]].
\bibitem{Davison:2013jba}
R.~A.~Davison,
Phys. Rev. D \textbf{88}, 086003 (2013)
doi:10.1103/PhysRevD.88.086003
[arXiv:1306.5792 [hep-th]].












\bibitem{Scheel:1994yr}
M.~A.~Scheel, S.~L.~Shapiro and S.~A.~Teukolsky,
Phys. Rev. D \textbf{51}, 4208-4235 (1995)
doi:10.1103/PhysRevD.51.4208
[arXiv:gr-qc/9411025 [gr-qc]].
\bibitem{Christodoulakis:2013xha}
T.~Christodoulakis, N.~Dimakis and P.~A.~Terzis,
J. Phys. A \textbf{47}, 095202 (2014)
doi:10.1088/1751-8113/47/9/095202
[arXiv:1304.4359 [gr-qc]].
\bibitem{Arkani-Hamed:2002bjr}
N.~Arkani-Hamed, H.~Georgi and M.~D.~Schwartz,
Annals Phys. \textbf{305}, 96-118 (2003)
doi:10.1016/S0003-4916(03)00068-X
[arXiv:hep-th/0210184 [hep-th]].





\bibitem{Capozziello:2005ku}
S.~Capozziello, V.~F.~Cardone and A.~Troisi,
Phys. Rev. D \textbf{71}, 043503 (2005)
doi:10.1103/PhysRevD.71.043503
[arXiv:astro-ph/0501426 [astro-ph]].


\bibitem{Misner:1964je}
C.~W.~Misner and D.~H.~Sharp,
Phys. Rev. \textbf{136}, B571-B576 (1964)
doi:10.1103/PhysRev.136.B571

\bibitem{Hayward:1997jp}
S.~A.~Hayward,
Class. Quant. Grav. \textbf{15}, 3147-3162 (1998)
doi:10.1088/0264-9381/15/10/017
[arXiv:gr-qc/9710089 [gr-qc]].





\bibitem{Padmanabhan:2002xm}
T.~Padmanabhan,
Mod. Phys. Lett. A \textbf{17}, 1147-1158 (2002)
doi:10.1142/S0217732302007260
[arXiv:hep-th/0205278 [hep-th]].
\bibitem{Wu:2008ir}
S.~F.~Wu, B.~Wang, G.~H.~Yang and P.~M.~Zhang,
Class. Quant. Grav. \textbf{25}, 235018 (2008)
doi:10.1088/0264-9381/25/23/235018
[arXiv:0801.2688 [hep-th]].


\bibitem{Tsilioukas:2024seh}
S.~A.~Tsilioukas, N.~Petropoulos and E.~N.~Saridakis,
Phys. Rev. D \textbf{111}, no.10, 103514 (2025)
doi:10.1103/PhysRevD.111.103514
[arXiv:2412.21146 [gr-qc]].


\bibitem{Bousso:1999xy}
R.~Bousso,
JHEP \textbf{07}, 004 (1999)
doi:10.1088/1126-6708/1999/07/004
[arXiv:hep-th/9905177 [hep-th]].
\bibitem{Gong:2006ma}
Y.~Gong, B.~Wang and A.~Wang,
JCAP \textbf{01}, 024 (2007)
doi:10.1088/1475-7516/2007/01/024
[arXiv:gr-qc/0610151 [gr-qc]].
\bibitem{Bamba:2009id}
K.~Bamba, C.~Q.~Geng and S.~Tsujikawa,
Phys. Lett. B \textbf{688}, 101-109 (2010)
doi:10.1016/j.physletb.2010.03.070
[arXiv:0909.2159 [gr-qc]].










\end{thebibliography}


\end{document}